# A NEW CONCEPT FOR SAFEGUARDING AND LABELING OF LONG-TERM STORED WASTE AND ITS PLACE IN THE SCOPE OF EXISTING TAGGING TECHNIQUES

Dina Chernikova[1], Kåre Axell[1,2]

[1]*Chalmers University of Technology, Department of Applied Physics, Nuclear Engineering, SE-412 96 Gothenburg, Sweden*

[2] *Swedish Radiation Safety Authority, SE-171 16 Stockholm, Sweden*

## Abstract

The idea of a novel labeling method is suggested for a new way of long-term security identification, inventory tracking, prevention of falsification and theft of waste casks, copper canisters, spent fuel containers, mercury containers, waste packages and other items. The suggested concept is based on the use of a unique combination of radioisotopes with different predictable half life. As an option for applying the radioisotope tag to spent fuel safeguarding it is suggested to use a mixture of α-emitting isotopes, such as $^{241}$Am etc., with materials that easily undergo α-induced reactions with emission of specific γ-lines. Thus, the existing problem of the disposing of smoke detectors or other devices [1] which contain radioisotopes can be addressed, indirectly solving an existing waste problem. The results of the first pilot experiments with two general designs of storage canisters, namely a steel container which corresponds to the one which is commonly used for long-term storing of mercury in Europe and USA and a copper canister, the one which is in applications for nuclear repositories, are presented. As one of the options for a new labeling method it is proposed to use a multidimensional bar code symbology and tungsten plate with ultrasound techniques. It is shown that the new radioisotope label offers several advantages in the scope of existing tagging techniques (overview is given) and can be implemented even with low activity sources.

## Introduction

"…mercury, alpha waste, high level waste (HLW), etc." – all these notions are related to the category of so-called long-term stored waste that are aimed at the disposition in geological repository [2]. There are number of classifications that exist for long-term stored waste. Generally, waste is separated into two groups: not radioactive, e.g. mercury waste, and radioactive that is coming from various parts of the nuclear fuel cycle, medical, industrial and research activities. However, there are uniting factors for all of them – long term management issues. One of these issues is related to sealing and containment verification technologies that can meet all needs for maintaining continuity of knowledge of waste in containers [3].

Various countries address this in different ways. For example, in US 10 CFR 60.135 regulations for HLW package design a unique identification is considered as one of the specific acceptance criteria:

> *"…(4)* Unique identification*. A label or other means of identification shall be provided for each waste package. The identification shall not impair the integrity of the waste package and shall be applied in such a way that the information shall be legible at least to the end of the period of retrievability. Each waste package identification shall be consistent with the waste package's permanent written records"* [4].

Similar criteria, i.e. "Qualitative acceptance criteria for radioactive wastes to be disposed of in deep geological formations" are discussed in Nirex Report (United Kingdom, UK):

> *"…*Unique identification*. Criteria. Each waste package for emplacement in a repository should be marked with a unique identification. Additional criteria. Records should be kept at different locations, nationally and internationally. Records should include information on location, chemical and physical properties of the waste; repository design and the information used for final safety assessment'* [5].

Thus, there is a noticeable trend towards the implementation of a unique labeling of spent fuel waste canisters which are aimed at the emplacement in a repository.

## Setting up requirements to the ideal tagging system

While choosing a particular type of tag it is necessary to consider a number of important parameters. There were a few attempts to systematize criteria for the selection of a specific tag, for example, based on: purpose of tag, type of the container, robustness, reliability, ease of application, effectiveness, interface with other safeguards and security elements, cost etc. [6].

Although, in connection with a long-term (hundreds of years) stored item, such as nuclear waste, spent fuel or mercury containers, one can consolidate these requirements in five main points ("intuitive requirements"), i.e. the ideal tag must provide:

1. Environmental safety (avoid corrosion effects of e.g. copper canisters). *The labeling system should avoid corrosion effects of canisters which can be induced in the long-term run, thus for instance avoiding leakage of spent fuel waste components later on.*
2. Non-contact reader system (preferably).

3. Long operation time. *The labeling system should have an operating time at least from ten to a few hundred years.*
4. Large and unique tag memory. *The labeling system should enable fully unique identification of the canister content in a manner consistent with permanent records of the storage or repository.*
5. Security technique against falsification of data, errors/multiple verification. *The labeling system should have high level of security, i.e. low risk of falsification or error.*

Thus, an ideal identification tag meets all the challenges of the international initiative on a holistic Safety, Security and Safeguards ('3S') concept. Therefore, hereafter we will consider the suitability of the currently existing technologies and new approach to these "intuitive requirements".

## Overview of existing methods

The conventional tagging techniques include etching characters, affixing identification plates, welding, etc. However, when considering an application for long-term storage of waste canisters they have a number of gaps in the factors of environmental safety, security and long operation time. Other disadvantages of the traditional labeling technology are described in [7]. Modern labeling techniques (Figure 1) may partly solve these problems and be useful for meeting the goals of a unique labeling system compatible with the record keeping of the storage or repository. Among the modern labeling techniques are radio frequency tagging systems, electronic tags, ultrasonic systems [8] and RPT (Reflective Particle Tags) [9], SANAs (SERS-Active Nanoparticle Aggregates; SERS: Surface Enhanced Raman Scattering) [10], etc.

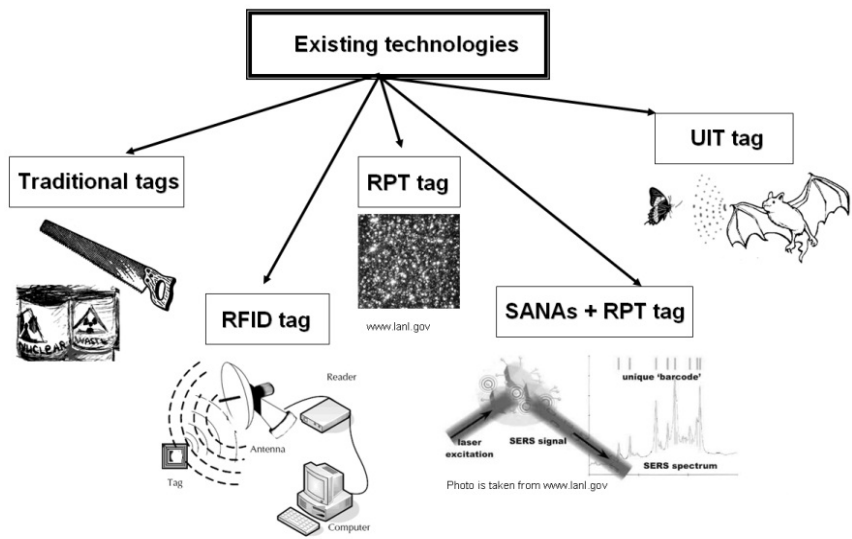

Figure 1. Existing tagging

The main disadvantages of these techniques are analyzed in [11]. Hereafter we only give a short overview of them in the light of the previously defined "intuitive requirements".

Radio frequency systems (RF) consist of a memory chip, an antenna, and a transmitter/receiver system and therefore overcome problems related to printing or etching characters on the side of the container. RF devices can be active, passive and semi-passive, as shown in Figure 2. Active tags contain a small internal power source to communicate, store and process large amounts of information in the chip. A power source is usually a lithium battery lasting less than 5 years. This makes them unsuitable for use in long-term storages. Passive tags have no battery. In order to provide power and data to the chip, they use the current in the loop antenna which is induced by the interrogating 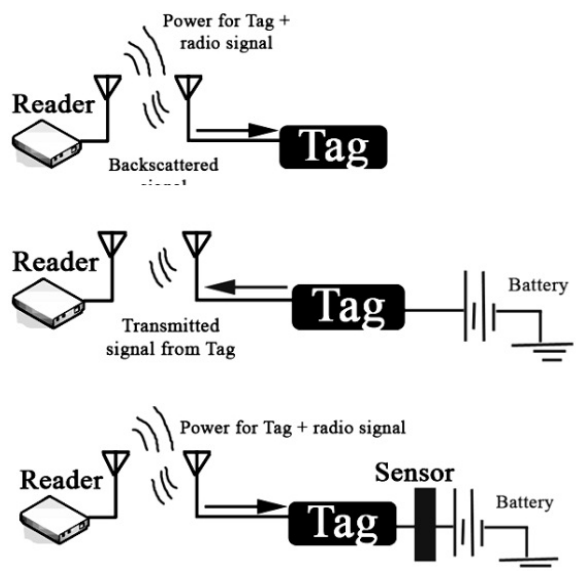 RF signal. Thus, they receive power from the reader's antenna. The main problems encountered with both active and passive RF devices is related to interference of the metallization layer with the RF signal, locating methods and low transmission range.

Ultrasonic tagging is based on the assumption of the uniqueness of the welding area of the cask. Thus, it assumes that in the process of ultrasonic scanning one can obtain a unique fingerprint for each stored container. However, it is difficult to explicitly evaluate performance of UIT in terms of environmental safety, long operating time and security. According to results of tests performed in [12] UIT methods suffer from problems with unknown long-term signature stability and material sensitivity, problems with repeatability of the signature and influence by the human factor.

Reflective Particle Tags (RPT) have been proposed by Sandia National Laboratories (SNL) in 1992. The tag represents the transparent adhesive matrix with encapsulated reflective particles. Although this system would be good enough to provide the identification for non-nuclear long-stored waste it will be difficult to apply it to casks containing radioactive material due to the difficulties connected to the reader system (a number of lights which induce the reflection in the tag) and presence of gamma background outside the cask walls. The characteristics of the reflective particle tag regarding long operating times, a large and unique tag memory and security can not be evaluated explicitly due to the present research stage of the technology. Results of tests performed in [12] indicate that the main drawbacks of RPT are related to the image degradation, inconsistent calibrations, occasional reader head instability and false rejection rate caused by corrosion. SANAs technology has unique strengths suited to a number of applications. However, due to the similar nature to RPT, it can suffer from the same problems as RPT techniques.

Accordingly there is a recognized need for a labeling system which last at least from ten to a few hundred years (time factor), at the same time enabling fully unique identification of the canister contents in a manner consistent with permanent records of the storage or repository (information factor), have a high level of security, i.e. low risk of falsification or error (security factor), and give the possibility to avoid a corrosion effect of canisters induced by the traditional tagging methods (environmental factor).

## Proposed approach and its application to the nuclear waste containers

The main idea of the proposed method consists of using a unique combination of radioisotopes with different predictable length of life and a long operating time, wherein the unique combination of radioisotopes comprises the mixture of two or more radioisotopes [13]. Radioisotopes have unique inherent properties, such as long half-life (hundreds of years), different penetration properties and energy characteristics (lines in the spectrum emitted by radioisotopes). These properties make them extremely attractive for use in tagging of long-term stored items, since they automatically provide: environmental safety (radioisotopic tag can be placed inside the canister), non-contact reader system, long operating time. The combination of radioisotopes should be chosen independently for each cask/waste container.

The majority of the background gamma rays in spent fuel originates from activation and fission products, e.g. $^{137}$Cs (662 keV (0.9) γ-line), $^{134}$Cs (569 keV (0.15), 605 keV (0.98), 796 keV (0.85), 802 keV (0.09), 1039 keV (0.01), 1168 keV (0.02) and 1365 keV (0.03) γ-lines), $^{144}$Pr (697 keV (0.0148), 1489 keV (0.003) and 2185 keV (0.008) γ-lines), $^{154}$Eu (723 keV (0.19), 873 keV (0.12), 996 keV (0.1), 1005 keV (0.17), 1275 keV (0.36) and 1595 keV (0.03) γ-lines) and $^{106}$Ru (512 keV (0.21), 622 keV (0.1), 1051 keV (0.02), 1128 keV (0.004) and 1357 keV (0.006) γ-lines). For a fuel cooled for a short period of time (less than four years), the high energy gamma lines, e.g. a 2185 keV gamma line from $^{144}$Pr, will be possible to measure. However, when the fuel will be sent to an encapsulation plant after a number of years of cooling, $^{137m}$Ba, the daughter nuclide of $^{137}$Cs, will be the main gamma emitter. Thus, if the radioisotope tag will have high energy signatures, there will be no problem with radiation background coming from the fuel.

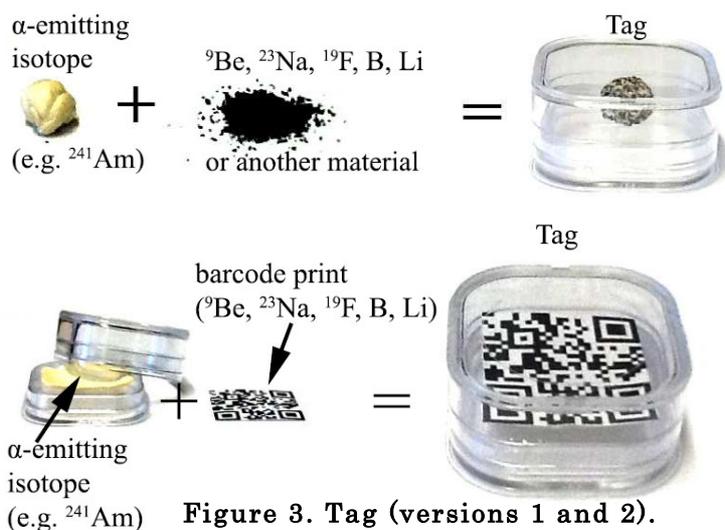

Figure 3. Tag (versions 1 and 2).

The simplest version of the conventional radioisotope tag may just include the specific radioisotopes which emits γ-rays with
energies higher than 1 MeV. Although, access to these isotopes can be restricted or their cost might be rather high. Therefore, we suggest to use the following version of the radioisotope tag based on α-emitting isotopes such as $^{241}$Am, for example, in a mixture with one of the materials described in [14], as shown in Figure 3. This version of the tag will serve the needs of long-term tagging of nuclear waste, as well as it can solve the existing problem of disposing of smoke detectors or other devices (surge voltage protection devices, electronic valves etc. [1]) which nowadays contain radioisotopes such as $^{241}$Am. It should be mentioned that according to the Report of the EU commission [1], as of the balance sheet date of year 2001, Ireland manufactured two million ionization chamber smoke detectors per year (activity of each detector is 33.3 - 37 kBq), while for example Sweden imported 700 000 of them. Thus, the price of the radioisotope tag based on this type of waste will be partly covered by the costs of the waste disposing. At the same time this method will open the possibility of recycling nuclear waste of this type.

One of the attractive options for realization of the radioisotope tag is implementation of the multidimensional bar code symbology, for example in a way as shown in Figures 3-4.

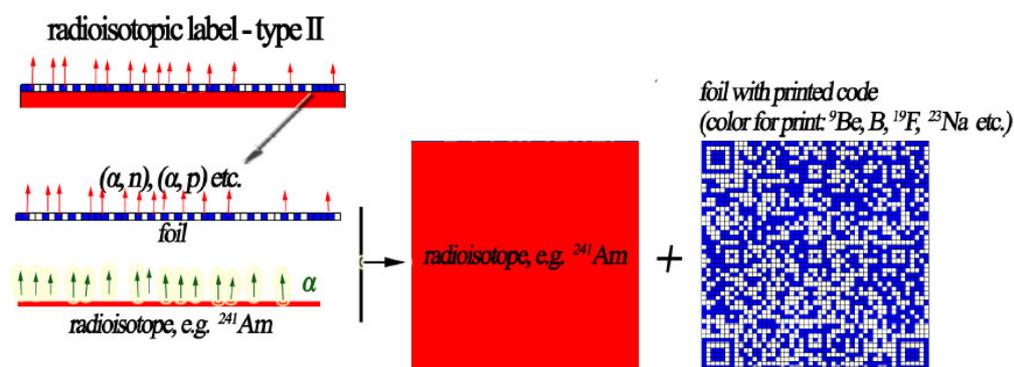

This version of the tag is suitable for a situation where the information about the item is not known in advance and should be encoded in the unique radioisotope tag directly at the encapsulation plant. This realization of the tag includes two components: a radioisotope plate prepared by authorities at a facility licensed for this work and a foil which is printed at the encapsulation plant. This concept is appealing in that while printing the tag a specific color could be used. As an example, the base of the tag can be made of an $^{241}$Am α-emitting isotope. Afterwards, the foil which contains the bar code could be printed with colors based on $^9$Be, $^{23}$Na, $^{19}$F, $^{10;11}$B, $^{30}$P, $^{7;6}$Li etc. materials. The printed foil must be placed in close contact with the α-emitting base of the tag. These materials have a high cross-section for α-induced reactions, such as (α,n), (α,p) etc. Thus, the bar code might be read detecting α-induced gamma rays. The energy of the gamma rays depends on the material which is chosen for printing the tag.

## Conclusions

We have described a new concept of a long-term security identification tags/labels which is based on the use of unique combinations of radioisotopes. In the case of application of this concept to spent fuel safeguarding it is suggested to use a mixture of α-emitting isotopes, such as $^{241}$Am with materials that easily undergo α-induced reactions with emission of specific γ-lines. Thus, if the radioisotope tag will have a high energy signature, there will be no problem with radiation background coming from the fuel. Moreover, this version of the radioisotope tag allows to solve the existing problem of the disposing of smoke detectors or other devices [1] which contain radioisotopes, such as $^{241}$Am, thus, indirectly providing a recycling of nuclear waste. As an economical advantage, it should be mentioned that the price of the radioisotope tag based on this type of waste will be partly covered by the costs of the waste disposing. As an attractive option for a new labeling method we proposed the possibility to realize a multidimensional bar code symbology. The new radioisotope label offers several advantages, as compared to the currently used tagging methods. It provides environmental safety, non-contact reader system, long operating time, large and unique tag memory, security technique against falsification of data, errors/multiple verification, recycling option for ionization chamber smoke detectors and other devices. Further details can be found in [14].

## Acknowledgement

This talk was supported by the Knut och Alice Wallenbergs Stiftelse "Jubileumsanslaget" 2013 C 2013/109, Sweden.